\documentclass[twocolumn,pre,superscriptaddress,amsmath,amssymb,longbibliography]{revtex4-1}
\usepackage{hyperref}
\usepackage{graphicx}
\usepackage{dcolumn}
\usepackage{bm}
\usepackage{float}
\usepackage{color}

\begin{document}


\title{Exact Expansion Formalism for Transport Properties of Heterogeneous Materials Characterized by Arbitrary Continuous Random Fields}

\author{Liyu Zhong}
\affiliation{Department of Mechanics and Engineering Science, College of Engineering, Peking University, Beijing 100871, P. R. China}
\author{Sheng Mao}
\email[correspondence sent to: ]{maosheng@pku.edu.cn}
\affiliation{Department of Mechanics and Engineering Science, College of Engineering, Peking University, Beijing 100871, P. R. China}



\date{\today}

\begin{abstract}
We derive an exact contrast‐expansion formalism for the effective conductivity of heterogeneous materials (media) with local properties described by arbitrary continuous random fields, significantly generalizing the widely used binary-field models. The theory produces a rapidly convergent Neumann‐series that, upon Gaussian closure via a Hermite expansion, yields closed‐form first-, second‐ and third‐order approximations, which achieve percent‐level accuracy at first order for isotropic media. For anisotropic media, second‐order approximations achieve sub-\(2\%\) accuracy across a wide range of local property contrasts and correlations. Our formalism provides mathematically rigorous structure–property closures, with significant implications for the discovery and design of novel graded and architected materials with tailored transport properties.
\end{abstract}

\maketitle

\section{Introduction}

Predicting effective properties of  heterogeneous materials (media) is a longstanding challenge in condensed matter physics and materials science \cite{To02a,Mi02,Sahimi2011}.  Composites, porous media, granular matter are examples of complex materials exhibiting emergent macroscopic properties that depend sensitively on their disordered microstructure. Classical analytical approaches, ranging from Maxwell’s mixing rule \cite{Maxwell1891} and Hashin–Shtrikman variational bounds \cite{hashin1963variational} to Beran’s weak‐contrast expansion \cite{Beran1965,Mi02}, have provided deep insight into the structure-property relationship of such material systems. 

Recently, data-driven approaches, particularly machine learning (ML) models, have been explored to establish microstructure–property mappings directly from experimental or synthetic data \cite{FENG2021110476,KAZEMIKHASRAGH2024113206,Xie2017,WEI2022100153, Cang2017, Bessa2017}. Although promising, many ML approaches operate as black-box predictors, lacking mathematical rigor nor physical interpretability \cite{cheng2022data}. As a consequence, their predictions are difficult to extrapolate beyond the training data \cite{Pokuri2019, Geirhos2020,HendrycksGimpel2017}. This lack of interpretability and limited generalizability pose obstacles for their adoption in new material design \cite{Guidotti2018,Samek2017,Arrieta2020}, where understanding \emph{why} a certain microstructure yields a given property is as important as the prediction itself \cite{Pokuri2019}. In this context, exact structure–property closures remain highly valuable—both as benchmarks and as foundations for computational design frameworks.

Notably, for statistically homogeneous two-phase (binary) media with arbitrary microstructures, a class of exact contrast expansion formalisms have been developed, which express the effective material properties as an infinite series involving integrals of the $n$-point correlation functions $S_n$ of material microstructure \cite{Beran1965, Mi81b, Mi02, torquato1985effective, torquato1997effective,torquato1997exact, Pham2003, Rechtsman2008, torquato2021nonlocal,Kim2020}. Pioneered by Beran \cite{Beran1965} and Milton \cite{Mi81b, Mi02}, and significantly generalized by Torquato \cite{torquato1985effective, torquato1997effective,torquato1997exact, Pham2003, Rechtsman2008, torquato2021nonlocal,Kim2020}, these expansions, based on carefully constructed dimensionless phase-property contrast parameters and when truncated at third- or fourth-order terms, yield accurate approximations of effective material properties in regimes of modest or strong phase contrast or moderate microstructural complexity.

\begin{figure}[H]
  \centering
  \includegraphics[width=0.45\textwidth]{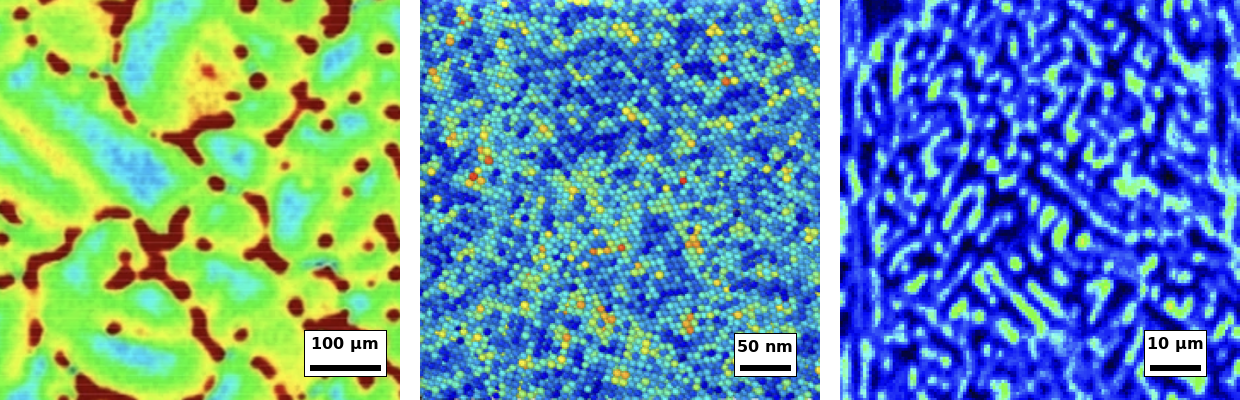}
  \caption{Selected examples highlighting continuous material heterogeneity.  
    Left panel: Mg distribution in an Al–Mg–Si–Cu–Mn alloy heat-treated at 300 °C for 8 h, showing compositional variations of Mg dispersoids \cite{Li2021Improved}.  
    Middle panel: Grain-interior (GI) segregation of Al in nanocrystalline NiCoAl from molecular-dynamics simulations, illustrating the effect of solute on dislocation interactions \cite{Zhang2023AlSegregation}.  
    Right panel: Fe distribution in an Al–Mn–Fe alloy EPMA map \cite{Song2017SoluteDistribution}.}
  \label{fig:real_micro_examples}
\end{figure}

Real-world materials always exhibit continuous heterogeneity rather than distinct phases, 
which may arise from diffusive interfaces in composites or continuous solute distributions in alloys (see examples in Fig.~\ref{fig:real_micro_examples}). Consequently, their local properties, such as conductivity or refractive index, vary smoothly in space \cite{Matheron1963,Dagan1982,Erdogan1995,Gelhar1992, ke2019enhanced, zheng2019modeling}. Continuous random field (CRF) models generalize classical two‑phase descriptions by allowing local material properties to take a continuum of values \cite{Wiersma2013, Dagan1989}, characterized by prescribed correlation functions, where existing contrast expansion formalisms cannot be applied.   

Here, we develop an exact contrast‐expansion formalism for continuous random fields that yields a fully tensorial infinite‐series representation of the effective conductivity. Starting from a cavity‐field integral equation, we derive a Neumann‐series whose terms involve connected cumulants of the local contrast field. Specializing to Gaussian statistics via a Hermite–Wick expansion produces explicit spectral integrals for the second‐ and third‐order approximations. Through extensive finite‐element simulations, we show that first-order approximations can achieve percent‐level accuracy for isotropic media; while for anisotropic media, second‐order approximations achieve sub-\(2\%\) accuracy across a wide range of local property contrasts and correlations.

\section{Method}
\label{sec:theory}

In this section we present the theoretical framework of our approach, which is structured in three parts. Firstly, we reformulate the heterogeneous‐conduction problem via a cavity‐field integral equation for an equivalent dipole in a reference medium, yielding a self‐consistent expression for the local field in terms of a Green’s tensor (analogous to Torquato’s two‐phase formulation \cite{To02a}). Secondly, we apply a Neumann–series inversion of this integral equation and identify the connected cluster contributions (two‐point and three‐point cumulants) to obtain a systematic expansion of the effective conductivity up to third order. Thirdly, assuming the underlying scalar field is Gaussian, we close the hierarchy by means of a Hermite–Wick expansion, which expresses the second‐ and third‐order corrections \(\mathbf{\Delta}_2\) and \(\mathbf{\Delta}_3\) in closed form via the covariance function \(C(\mathbf{r})\). The resulting formulas provide explicit, physically transparent predictions for the full tensor \(\mathbf{\sigma}_e\).
\subsection{Cavity Reformulation via Dipole Integral Equation}

To avoid the intractability of directly solving the spatially varying–coefficient PDE for heterogeneous conduction, we reformulate the problem into an integral equation by embedding the medium in a uniform reference phase.  This Lippmann–Schwinger approach treats local conductivity deviations as dipole sources acting within the reference medium.

Consider a continuous scalar conductivity field \(\sigma(\mathbf{x})>0\) in \(\mathbb{R}^d\) under an imposed uniform far–field electric field \(\mathbf{E}_0\).  The local current density obeys Ohm’s law,
\begin{equation}
  \mathbf{J}(\mathbf{x}) = \sigma(\mathbf{x})\,\mathbf{E}(\mathbf{x}),
\end{equation}
together with the continuity condition
\begin{equation}
  \nabla\!\cdot\!\mathbf{J}(\mathbf{x}) = 0.
\end{equation}

\textcolor{red}{Embedding the medium in an infinite reference phase of uniform conductivity \(\sigma_q>0\) and defining the polarization field
\begin{equation}\label{eq:P_def}
  \mathbf{P}(\mathbf{x})
  = [\,\sigma(\mathbf{x}) - \sigma_q\,]\,\mathbf{E}(\mathbf{x}),
\end{equation}}
allows us to write the total current as \(\mathbf{J} = \sigma_q\,\mathbf{E} + \mathbf{P}\).  \textcolor{red}{Substitution into the Green’s–tensor representation of the reference medium yields the Lippmann–Schwinger integral equation \cite{To02a}:
\begin{equation}\label{eq:LS_integral}
  E_i(\mathbf{x})
  = E_{0,i}
  + \int_{\mathbb{R}^d} G^{(q)}_{ij}(\mathbf{x}-\mathbf{x}')\,P_j(\mathbf{x}')\,\mathrm{d}^d\mathbf{x}'.
\end{equation}}

Here, the Green’s tensor decomposes into a singular Lorentz term, which accounts for the self–interaction, and a smooth dipole kernel:
\begin{equation}\label{eq:G_decomp}
  G^{(q)}_{ij}(\mathbf{r})
  = -D^{(q)}_{ij}\,\delta(\mathbf{r})
    + H^{(q)}_{ij}(\mathbf{r}),
  \quad
  D^{(q)}_{ij} = \frac{1}{d\,\sigma_q}\,\delta_{ij},
\end{equation}
\begin{equation}
  H^{(q)}_{ij}(\mathbf{r})
  = \frac{1}{\Omega\,\sigma_q}\,\frac{d\,n_i n_j - \delta_{ij}}{r^d},
  \quad
  \mathbf{n} = \frac{\mathbf{r}}{r},
  \quad
  \Omega = \frac{2\pi^{d/2}}{\Gamma(d/2)}.
\end{equation}

Excising the singular term removes the local self–dipole contribution and leads to the definition of the cavity field:
\begin{equation}\label{eq:F_def}
  \mathbf{F}(\mathbf{x})
  = \mathbf{E}_0
  + \int_{\mathbb{R}^d} H^{(q)}_{ij}(\mathbf{x}-\mathbf{x}')\,P_j(\mathbf{x}')\,\mathrm{d}^d\mathbf{x}'.
\end{equation}

By comparing \eqref{eq:G_decomp} and \eqref{eq:LS_integral}, one sees that
\begin{equation}\label{eq:E_F_relation}
\mathbf{E}(\mathbf{x})=\mathbf{F}(\mathbf{x})-\mathbf D^{(q)}\!\cdot\!\mathbf P(\mathbf{x})\
\end{equation}

\textcolor{red}{which rearranges to a local relation for the polarization:
\begin{equation}\label{eq:P_local}
  \mathbf{P}(\mathbf{x})
  = \mathbf{L}^{(q)}(\mathbf{x})\cdot\mathbf{F}(\mathbf{x}),
  \quad
  \mathbf{L}^{(q)}(\mathbf{x}) = L^{(q)}(\mathbf{x})\,\mathbf{I},
\end{equation}
where \(\mathbf{I}\) denotes the \(d\times d\) identity tensor}, and
\begin{equation}
  L^{(q)}(\mathbf{x})
  = \bigl[\sigma(\mathbf{x})-\sigma_q\bigr]
    \Bigl[1 + \tfrac{\sigma(\mathbf{x})-\sigma_q}{d\,\sigma_q}\Bigr]^{-1}.
\end{equation}

Introducing the dimensionless contrast
\textcolor{red}{\begin{equation}\label{eq:beta_def}
  \beta(\mathbf{x})
  = \frac{\sigma(\mathbf{x})-\sigma_q}
         {\sigma(\mathbf{x}) + (d-1)\,\sigma_q},
  \quad
  L^{(q)} = \sigma_q\,d\,\beta,
\end{equation}}
we see that the polarization field \(\mathbf{P}(\mathbf{x})\) encapsulates all microscale fluctuations as dipole sources, while the cavity field \(\mathbf{F}(\mathbf{x})\) represents the smooth driving field free of local self–interactions.  This reformulation mirrors Torquato’s strong–contrast method for two–phase composites \cite{To02a} and sets the stage for the Neumann–series expansion of the effective conductivity tensor in the next section.

\subsection{Second- and Third-Order Connected Expansion for General Random Fields}

Building on the cavity-field integral formulation, we perform a Neumann-series expansion to extract the connected two-point and three-point contributions to the effective polarizability operator.  Here \(\mathbf H\) denotes the cavity-field dipole–kernel operator of \eqref{eq:G_decomp} and \(L\) the local polarizability operator of \eqref{eq:P_local}.

The operator equations
\begin{equation}
  \mathbf F = \mathbf E_0 + \mathbf H\,\mathbf P,
  \quad
  \mathbf P = L\,\mathbf F,
\end{equation}
imply
\begin{equation}
  \mathbf P = L\,\mathbf E_0 + L\,\mathbf H\,\mathbf P.
\end{equation}
Formally inverting by Neumann series,
\begin{equation}
  \mathbf P
  = L\,\bigl[\mathbf I - L\,\mathbf H\bigr]^{-1}\mathbf E_0
  = \mathbf S\,\mathbf E_0,
\end{equation}
and ensemble-averaging yields
\begin{equation}
  \langle\mathbf P\rangle = \langle \mathbf S\rangle\,\mathbf E_0,
\end{equation}
\begin{equation}
  \langle\mathbf F\rangle
  = \bigl[\langle\mathbf S\rangle^{-1} + \mathbf H\bigr]\,
    \langle\mathbf P\rangle
  = \mathbf Q\,\langle\mathbf P\rangle,
\end{equation}
so that
\begin{equation}
  \mathbf L_e^{(q)} = \mathbf Q^{-1}.
\end{equation}
The effective conductivity tensor \(\mathbf\sigma_e\) then follows from the fractional transform
\begin{equation}
  \mathbf L_e^{(q)}
  = \sigma_q\,d\;
    \{\mathbf\sigma_e - \sigma_q\,\mathbf I\}\;
    \{\mathbf\sigma_e + (d-1)\,\sigma_q\,\mathbf I\}^{-1}.
\end{equation}

Introduce the mean polarizability \(a=\langle L\rangle\) and fluctuation \(\delta L=L-a\), and define the connected cumulants,
\begin{equation}
  \mathbf\Delta_2
  = \bigl\langle \delta L(1)\,\mathbf H(1,2)\,\delta L(2)\bigr\rangle_{\mathrm{conn}},
  \quad
\end{equation}
\begin{equation}
  \mathbf\Delta_3
  = \bigl\langle \delta L(1)\,\mathbf H(1,2)\,\delta L(2)\,
    \mathbf H(2,3)\,\delta L(3)\bigr\rangle_{\mathrm{conn}},
  \quad
\end{equation}
where
$
\delta L(\mathbf x)
= d\,\sigma_q\bigl[\beta(g(\mathbf x))-\langle\beta\rangle\bigr]
=\;d\,\sigma_q\,\beta_c(\mathbf x),$
and expand \(\langle\mathbf S\rangle\) to third order in \(\mathbf H\).  Since we wish to retain all contributions through \(\mathcal O(H^2)\) in \(\langle\mathbf S\rangle^{-1}\) (so that \(\mathbf\Delta_3\) appears without additional mixing), we write
\begin{equation}
\begin{split}
  \langle\mathbf S\rangle
  &= a\,\mathbf I
   + a^2\,\mathbf H
   + a^3\,\mathbf H^2
   + \mathbf\Delta_2 \\[-3pt]
  &\quad + a\bigl(\mathbf\Delta_2\,\mathbf H + \mathbf H\,\mathbf\Delta_2\bigr)
   + \mathbf\Delta_3
   + \mathcal O(H^3),
\end{split}
\end{equation}
\begin{equation}
\begin{split}
  \langle\mathbf S\rangle^{-1}
  &= \frac{1}{a}\,\mathbf I
   - \mathbf H
   - \frac{\mathbf\Delta_2}{a^2}
   - \frac{\mathbf\Delta_3}{a^2} \\[-3pt]
  &\quad + \frac{\mathbf\Delta_2^2}{a^3}
   + \mathcal O(H^3).
\end{split}
\end{equation}
Adding back \(\mathbf H\) gives
\begin{equation}
\begin{split}
  \mathbf Q
  &= \langle\mathbf S\rangle^{-1} + \mathbf H \\[-3pt]
  &= \frac{1}{a}\,\mathbf I
    - \frac{\mathbf\Delta_2}{a^2}
    - \frac{\mathbf\Delta_3}{a^2}
    + \frac{\mathbf\Delta_2^2}{a^3}
    + \mathcal O(H^3),
\end{split}
\end{equation}
\begin{equation}
  \mathbf L_e^{(q)}
  = \mathbf Q^{-1}
  = a\,\mathbf I
    + \mathbf\Delta_2
    + \mathbf\Delta_3
    + \mathcal O(H^3).
\end{equation}

Defining
\begin{equation}
  \mathbf M
  = \frac{1}{\sigma_q\,d}\,\mathbf L_e^{(q)},
\end{equation}
the effective conductivity takes the form
\begin{equation}
  \mathbf{\sigma_e}
  = \sigma_q\,(\mathbf I + (d-1)\mathbf M)\,(\mathbf I - \mathbf M)^{-1},
\end{equation}
\begin{equation}
  \mathbf M
  = \langle\beta\rangle\,\mathbf I
    + \frac{\mathbf\Delta_2}{d\,\sigma_q}
    + \frac{\mathbf\Delta_3}{d\,\sigma_q}
    + \mathcal O(H^3).
\end{equation}

The real-space forms of the cumulants are
\begin{equation}\label{eq:Delta2}
  (\Delta_2)_{ij}
  = d^2\,\sigma_q
    \int d^d\mathbf r\;
    H^{(q)}_{ij}(\mathbf r)\;
    \langle\beta_c(0)\,\beta_c(\mathbf r)\rangle,
\end{equation}
\begin{equation}\label{eq:Delta3}
\begin{split}
  (\Delta_3)_{ij}
  &= d^3\,\sigma_q
     \iint d^d\mathbf r_1\,d^d\mathbf r_2 \\[-3pt]
  &\quad \times
     H^{(q)}_{ik}(\mathbf r_1)\,
     H^{(q)}_{kj}(\mathbf r_2)\,
     \langle\beta_c(0)\,\beta_c(\mathbf r_1)\,\beta_c(\mathbf r_2)\rangle.
\end{split}
\end{equation}

Having obtained the connected corrections \(\mathbf\Delta_2\) and \(\mathbf\Delta_3\) for an arbitrary continuous random field, the next section will employ the Gaussian–field Hermite–Wick expansion to close these expressions and derive explicit spectral-domain formulas.

\subsection{Gaussian‐Field Closure via Hermite–Wick Expansion}

In order to close the hierarchy of cumulant equations \eqref{eq:Delta2} and \eqref{eq:Delta3} for a continuous Gaussian random field, we introduce a zero–mean, unit–variance Gaussian field \(g(\mathbf x)\) with two–point covariance
\begin{equation}
\langle g(\mathbf x)\,g(\mathbf x')\rangle
= C\bigl(|\mathbf x-\mathbf x'|\bigr)\,,
\end{equation}
and recover the scalar conductivity via the affine mapping
\begin{equation}
\sigma(\mathbf x) = m + s\,g(\mathbf x)\,,
\end{equation}
where \(m\ge3s>0\) ensures \(\sigma(\mathbf x)>0\) with overwhelming probability.  The associated bounded polarizability field
\begin{equation}
\beta(g)
= \frac{\sigma(\mathbf x)-\sigma_q}{\sigma(\mathbf x)+(d-1)\,\sigma_q}
= \frac{A + s\,g}{B + s\,g}
\end{equation}
\begin{equation}
A = m - \sigma_q,\quad
B = m + (d-1)\,\sigma_q.
\end{equation}
is expanded in probabilists’ Hermite polynomials \(H_k(g)\) with coefficients
\begin{equation}
\beta(g)
= \sum_{k=0}^{\infty} a_k\,H_k(g)
\end{equation}
\begin{equation}
a_k
= \frac{1}{k!}
  \int_{-\infty}^{\infty}
    \beta(u)\,H_k(u)\,\phi(u)\,du
\end{equation}
\begin{equation}
\phi(u)
= \frac{e^{-u^2/2}}{\sqrt{2\pi}}
\end{equation}
Wick’s theorem for Gaussian fields then yields closed‐form expressions for the connected two‐ and three‐point cumulants of \(\beta\).  Truncating at \(k\le3\) gives
\begin{equation}
\begin{split}
\langle\beta_c(0)\,\beta_c(\mathbf r)\rangle
&= \sum_{k=1}^3 k!\,a_k^2\,[C(r)]^k\\
&= a_1^2\,C(r)
  + 2\,a_2^2\,C(r)^2
  + 6\,a_3^2\,C(r)^3
\end{split}
\end{equation}
\begin{equation}
\begin{split}
\langle\beta_c(0)\,\beta_c(\mathbf r_1)\,\beta_c(\mathbf r_2)\rangle
&=2\,a_1^2a_2\Bigl[C(r_1)C(r_2)\\
 &\quad+C(r_1)C(r_{12})
 +C(r_2)C(r_{12})\Bigr]\\
&\quad+8\,a_2^3\,C(r_1)\,C(r_2)\,C(r_{12})
\end{split}
\end{equation}

where \(r=|\mathbf r|\), \(r_i=|\mathbf r_i|\), and \(r_{12}=|\mathbf r_1-\mathbf r_2|\).  Substitution into the real–space cumulant integrals \eqref{eq:Delta2} and \eqref{eq:Delta3} then yields the Hermite–Wick closure for the second– and third–order corrections:
\begin{equation}
\begin{split}
(\Delta_2)_{ij}
&= d^2\,\sigma_q
  \int d^d\mathbf r\,
  H^{(q)}_{ij}(\mathbf r)\\
&\quad\times \Bigl[
    a_1^2\,C(r)
    + 2\,a_2^2\,C(r)^2
    + 6\,a_3^2\,C(r)^3
  \Bigr]
\end{split}
\end{equation}
\begin{equation}
\begin{split}
(\Delta_3)_{ij}
&= d^3\,\sigma_q
  \iint d^d\mathbf r_1\,d^d\mathbf r_2\,
  H^{(q)}_{ik}(\mathbf r_1)\,H^{(q)}_{kj}(\mathbf r_2)\\
&\quad\times \Bigl\{
    2\,a_1^2a_2\Bigl[C(r_1)C(r_2)\\
 &\quad+C(r_1)C(r_{12})
 +C(r_2)C(r_{12})\Bigr]\\
&\quad+8\,a_2^3\,C(r_1)\,C(r_2)\,C(r_{12})
  \Bigr\}
\end{split}
\end{equation}

Finally, we transform Eqs.~\eqref{eq:Delta2} and \eqref{eq:Delta3} into the Fourier domain.  Defining the spectra
\begin{equation}
\widetilde C(q)
= \int C(r)\,e^{-i\mathbf q\cdot\mathbf r}\,d^d\mathbf r,
\end{equation}
\begin{equation}
\widetilde{C^n}(q)
= \int [C(r)]^n\,e^{-i\mathbf q\cdot\mathbf r}\,d^d\mathbf r,
\end{equation}
and the dipole‐kernel tensor
\begin{equation}
\Gamma_{ij}(\mathbf q)
= \frac{q_i q_j}{q^2} - \frac{\delta_{ij}}{d},
\end{equation}
the second‐order correction becomes
\begin{equation}
\begin{split}
(\Delta_2)_{ij}
&= d^2\,\sigma_q
  \int\frac{d^d\mathbf q}{(2\pi)^d}\,
  \Gamma_{ij}(\mathbf q)\\
&\quad\times \Bigl[
    a_1^2\,\widetilde C(q)
    + 2\,a_2^2\,\widetilde{C^2}(q)
    + 6\,a_3^2\,\widetilde{C^3}(q)
  \Bigr]
\end{split}
\end{equation}
For the third‐order term, we introduce the three‐point spectrum
\begin{equation}
\begin{split}
\widetilde C^{(3)}_{\beta}(\mathbf q,\mathbf p)
&=2\,a_1^2a_2\Bigl[
    \widetilde C(q)\,\widetilde C(p)
  + \widetilde C(q)\,\widetilde C(|\mathbf q+\mathbf p|)\\
&\quad
  + \widetilde C(p)\,\widetilde C(|\mathbf q+\mathbf p|)
\Bigr]\\
&\quad+8\,a_2^3\,
  \widetilde C(q)\,\widetilde C(p)\,\widetilde C(|\mathbf q+\mathbf p|)
\end{split}
\end{equation}
In the macroscopically uniform limit \(\mathbf k=0\) one obtains
\begin{equation}
\begin{split}
(\Delta_3)_{ij}
&= d^3\,\sigma_q
  \iint\frac{d^d\mathbf q}{(2\pi)^d}\frac{d^d\mathbf p}{(2\pi)^d}\,
  \Gamma_{ik}(\mathbf q)\,\Gamma_{kj}(\mathbf p)\,\\&\quad\times(2\pi)^d\delta(\mathbf q+\mathbf p)
 \widetilde C^{(3)}_\beta(\mathbf q,\mathbf p),
\end{split}
\end{equation}
which reduces to
\begin{equation}
(\Delta_3)_{ij}
= d^3\,\sigma_q
  \int\frac{d^d\mathbf q}{(2\pi)^d}\,
  \Gamma_{ik}(\mathbf q)\,\Gamma_{kj}(\mathbf q)\,
  \widetilde C^{(3)}_\beta(\mathbf q,-\mathbf q).
\end{equation}
Substitution of these spectral corrections into the fractional transform then yields the explicit Fourier‐domain representation of the effective conductivity tensor.
\section{Microstructure Reconstruction and Spectral Design of Gaussian Random Fields}
\label{sec:microstructure}

\begin{figure*}[htbp]
  \centering
  \includegraphics[width=0.9\textwidth]{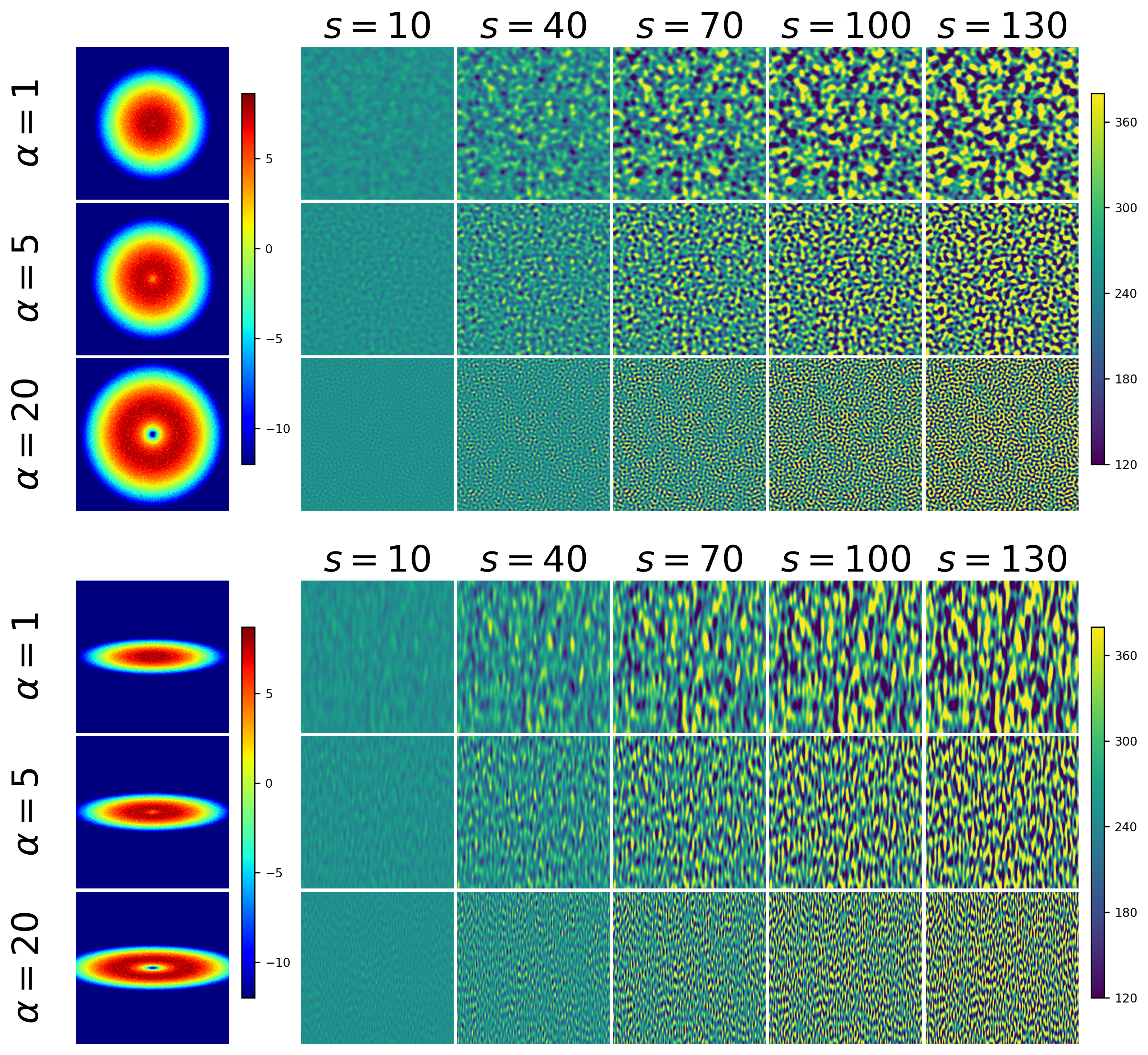}
  \caption{Representative realizations of Gaussian random fields used to model the local conductivity of heterogeneous media.  The upper block (\emph{rows labelled $\alpha=1,5,20$}) displays isotropic microstructures; the lower block shows the corresponding anisotropic cases with aspect ratio $a_y/a_x=4$.  In each block, the leftmost panel visualises the logarithmic power-spectral density $S_g(k_x,k_y)$, while the five panels to the right depict the conductivity fields $\sigma=250+s\,g$ for contrast amplitudes $s=10,40,70,100,130$ (left to right).  A common linear colour scale $[250-3s,\,250+3s]$ is used within every row.  Increasing $\alpha$ progressively suppresses long-wavelength modes, leading to finer textures, whereas increasing $s$ simply widens the dynamic range of $\sigma$ without altering the spatial pattern.}
  \label{fig:combined_microstructures}
\end{figure*}

Our cumulant expansion in Sec.~\ref{sec:theory} produces closed-form second- and third-order corrections, $\bm{\Delta}_2$ and $\bm{\Delta}_3$, that depend only on the connected moments of the conductivity field.  To obtain explicit expressions we restrict attention to zero–mean, unit-variance Gaussian random fields and invoke the Hermite–Wick theorem, thereby expressing every cumulant in terms of the two-point covariance $C(\mathbf r)$.  Because $C(\mathbf r)$ is the Fourier transform of the power-spectral density (PSD) $S_g(\mathbf k)$, designing $S_g$ gives us full control over the microstructure’s second-order statistics.

\subsection{Spectral model and Generative Method}
Microstructures are generated by filtering white noise in Fourier space.  
Let $w(\mathbf x)$ be a discrete white-noise field on an $N\times N$ grid of spacing $\Delta x=L/N$.  
Its discrete Fourier transform $\hat w(\mathbf k)$ fulfils Hermitian symmetry so that the inverse transform is real.  
We impose a prescribed PSD by multiplying $\hat w(\mathbf k)$ with the square root of a target spectrum.

We begin with the unnormalised spectrum
\begin{equation}
  S_{\alpha}^{\mathrm{raw}}(\mathbf k)
  = \bigl(k_{\mathrm{eff}}+k_0\bigr)^{\alpha}
    \exp\!\Bigl[-\tfrac{k_{\mathrm{eff}}^{2}}{2\sigma_k^{2}}\Bigr],
  \label{eq:Salpha_raw}
\end{equation}
where $k_0\ll1$ regularises the origin and $\sigma_k$ damps high-frequency modes.  
The effective wavenumber
\begin{equation}
  k_{\mathrm{eff}}
  = 
  \begin{cases}
    \|\mathbf k\|\;, & \text{isotropic},\\[4pt]
    \sqrt{(k_x/a_x)^2 + (k_y/a_y)^2}\;, & \text{anisotropic},
  \end{cases}
  \label{eq:keff}
\end{equation}
introduces ellipticity when $a_x\neq a_y$.  
Larger $\alpha$ deepens the “notch’’ at low~$k$, suppressing long-wavelength fluctuations.

Because Eq.~\eqref{eq:Salpha_raw} does not normalise the variance, we divide by the total spectral power
\begin{equation}
  N_\alpha
  = \int_{\mathbb R^2}\!\frac{S_{\alpha}^{\mathrm{raw}}(\mathbf k)}{(2\pi)^2}\,d^2k,
  \qquad
  S_g(\mathbf k)=\frac{S_{\alpha}^{\mathrm{raw}}(\mathbf k)}{N_\alpha},
  \label{eq:normalisation}
\end{equation}
and set
\begin{equation}
  \hat g(\mathbf k)=\sqrt{S_g(\mathbf k)}\,\hat w(\mathbf k),
  \qquad
  g(\mathbf x)=\mathcal F^{-1}\!\{\hat g(\mathbf k)\}.
  \label{eq:filtering}
\end{equation}
Finally we subtract the sample mean and divide by the sample standard deviation so that $\langle g\rangle=0$ and $\operatorname{Var}[g]=1$ to machine precision.  The local conductivity field then follows from the affine map
\begin{equation}
  \sigma(\mathbf x)=m+s\,g(\mathbf x),
  \qquad
  m=250,
  \label{eq:sigma_map}
\end{equation}
with $s$ controlling the contrast amplitude.  
This Fourier-filtering procedure reproduces the target PSD to numerical accuracy and is widely used for simulating correlated Gaussian media\cite{shinozuka1991simulation,woodchan1994,ZHONG2025acta}.

\begin{figure*}[htbp]
  \centering
  \includegraphics[width=0.9\textwidth]{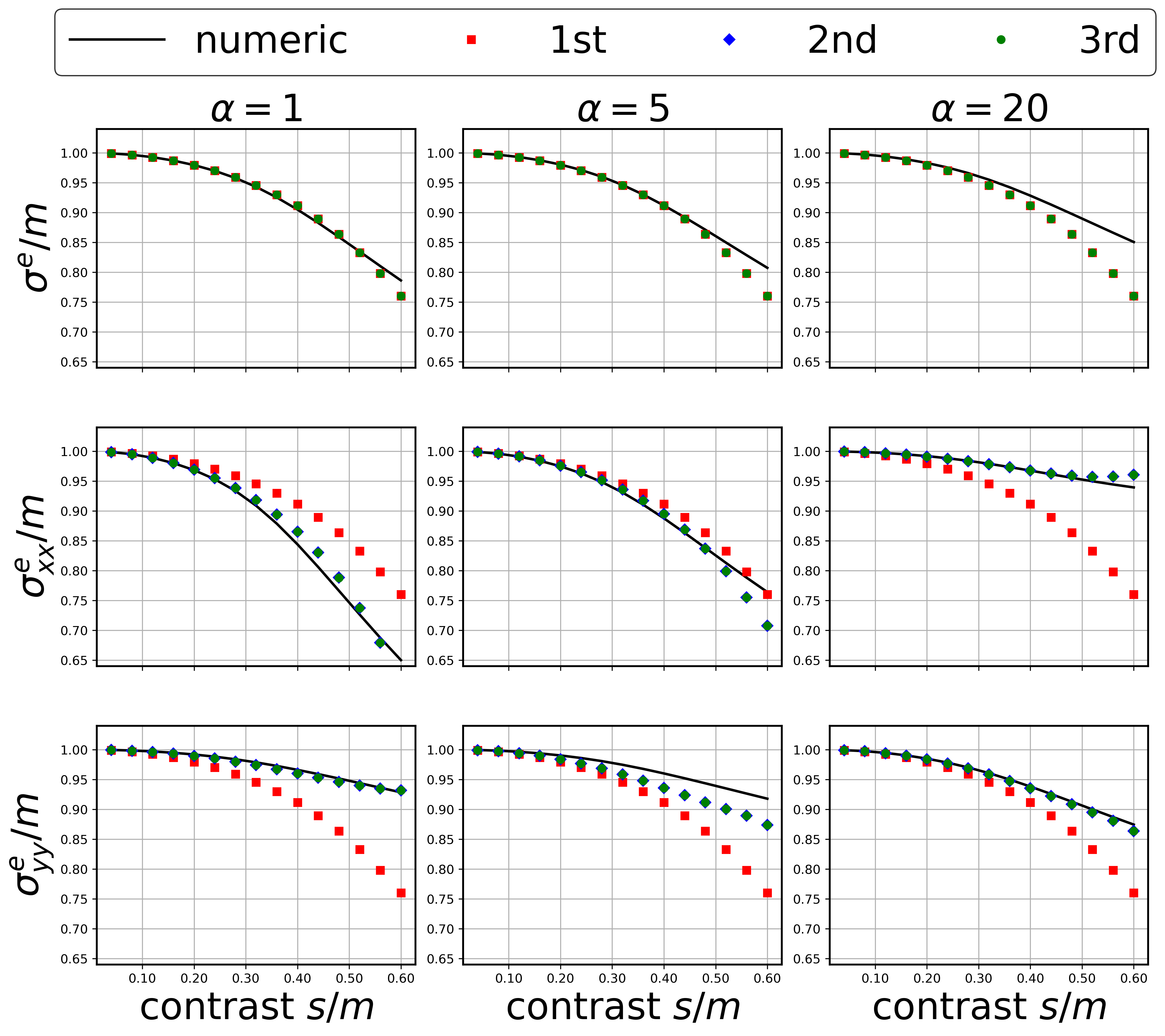}
  \caption{Effective conductivity of isotropic and anisotropic Gaussian media as a function of normalised contrast $\tilde{s}=s/m$ for spectral exponents $\alpha=1,5,20$ (columns).  The top row shows the scalar effective conductivity $\sigma^e/m$ for isotropic spectra.  The middle and bottom rows display the tensor components $\sigma^e_{xx}/m$ and $\sigma^e_{yy}/m$, respectively, for anisotropic spectra with aspect ratio $a_y/a_x=4$.  Numerical homogenisation results appear as black lines, while first-, second-, and third-order theoretical approximations are indicated by red squares, blue diamonds, and green circles.}
  \label{fig:combined_orders}
\end{figure*}

\subsection{Representative Microstructures}
Figure~\ref{fig:combined_microstructures} gathers a comprehensive set of realisations spanning both isotropic and anisotropic statistics.  Within each block the leftmost panel shows $S_g(\mathbf k)$ on a logarithmic scale.  For isotropic fields (top) the level sets are circular; for anisotropic fields (bottom) they are elliptical with aspect ratio $a_y/a_x=4$.  

Moving rightwards, the conductivity fields $\sigma(\mathbf x)=250+s\,g(\mathbf x)$ are displayed for five contrast levels $s=10,40,70,100,130$.  A single linear colour bar $[250-3s,\,250+3s]$ is used within each column so that changes in visual intensity reflect only the spatial structure of $g$.  

Three key trends emerge. Increasing $\alpha$ from~1 to~20 increasingly suppresses long-wavelength components, fragmenting broad patches into fine “grains’’. In the anisotropic block elongated level sets in Fourier space translate into vertically aligned streaks in real space, confirming the direct link between $S_g$ and spatial anisotropy. Larger $s$ does not alter spatial patterns but stretches the conductivity range linearly.  

These microstructures form the data set for the effective-property calculations reported in Sec.~\ref{sec:results}.  By systematically varying $\alpha$, $a_y/a_x$, and $s$ we sample a wide portion of parameter space, enabling stringent tests of the perturbative theory developed earlier.

\section{Results and Validation}
\label{sec:results}

We benchmark the analytical predictions developed in Secs.~\ref{sec:theory}–\ref{sec:microstructure} against large-scale numerical homogenisation.  The local conductivity field is prescribed as
\begin{equation}
  \sigma(\mathbf x)=m+s\,g(\mathbf x),\qquad m=250,
\end{equation}
so that the scalar $m$ represents the background matrix phase and the parameter $s$ controls the amplitude of fluctuations via the zero-mean, unit-variance Gaussian field $g(\mathbf x)$.  Three spectral exponents, $\alpha=1,5,20$, are examined, and the dimensionless contrast $\tilde{s}=s/m$ is varied from $0.04$ to $0.60$ (i.e.\ $s\in[10,\,150]$).  For every $(\alpha,\tilde{s})$ pair we generate $8\times10^{3}$ statistically independent realisations on a $256\times256$ grid.  Each realisation is solved with a high-order finite-element scheme under periodic boundary conditions, yielding the sample-wise effective tensor $\bm{\sigma}^e$.  Ensemble averaging then furnishes the numerical reference values.

For isotropic power spectra the effective tensor reduces to a single scalar, $\sigma^e=\tfrac{1}{d}\operatorname{tr}\bm{\sigma}^e$.  The dipole kernel $\Gamma_{ij}(\mathbf k)$ integrates to zero in the second-order cumulant, rendering $\bm{\Delta}_2$ identically vanishing and leaving the third-order term as the leading microstructural correction.  The third-order contribution is, however, numerically two orders of magnitude smaller than the first-order term, so the spectrum-independent first-order expression dominates.  The top row of Fig.~\ref{fig:combined_orders} shows that this first-order approximation tracks the numerical benchmark to within $2\%$ for $\tilde{s}\le0.40$ and remains inside a $10\%$ envelope even at the largest contrasts studied.  The third-order curve is practically indistinguishable from the first-order one, confirming that higher-order effects are negligible in isotropic media throughout the considered parameter space.

When the spectrum is anisotropic—here with aspect ratio $a_y/a_x=4$—the effective tensor exhibits two principal components, $\sigma^e_{xx}$ and $\sigma^e_{yy}$.  Because the first-order theory lacks spectral information, it cannot capture directional effects and therefore deviates noticeably from the numerical data, as evident from the red symbols in the middle and bottom rows of Fig.~\ref{fig:combined_orders}.  Including the second-order cumulant introduces the contracted product $\Gamma_{ij}(\mathbf k)S_g(\mathbf k)$ and restores the missing spectral content, bringing the prediction within $5\%$ of the numerical benchmark for all $\tilde{s}\le0.50$ and keeping the error below $10\%$ up to $\tilde{s}=0.60$.  The third-order correction adds only a minor refinement, indicating that the series is effectively converged at second order.  These findings demonstrate that the connected two-point cumulant captures the dominant anisotropic influence of the spectrum, whereas three-point effects play a secondary role.

Overall, Fig.~\ref{fig:combined_orders} confirms that the Hermite–Wick expansion delivers a rapidly convergent and quantitatively accurate description of effective transport in Gaussian random media.  For isotropic structures a spectrum-independent first-order formula suffices, while for anisotropic structures explicit inclusion of the two-point spectrum at second order narrows the error to a few percent even under strong contrast and strongly suppressed long-wavelength content.  These results validate the analytical framework advanced in this work and underscore its utility for predicting transport properties of complex disordered materials.

\section{Conclusions and Discussion}
This study has presented an exact contrast–expansion formalism for
predicting the effective conductivity of heterogeneous media whose local
properties are described by continuous random fields.
Starting from the cavity–field integral equation, we performed a
Neumann–series inversion and closed the resulting hierarchy with a
Hermite–Wick procedure that is exact for Gaussian statistics.
The resulting expressions yield closed‐form first-, second-, and
third-order corrections in terms of explicit spectral integrals,
thereby establishing a transparent connection between the
microstructure’s power‐spectral density and its macroscopic transport
response.

Comprehensive numerical homogenisation on more than
$10^{5}$ realisations confirmed the accuracy and rapid convergence of
the theory.  For isotropic spectra the second-order cumulant vanishes by
symmetry, and the spectrum‐independent first‐order term alone predicts
the effective conductivity within two per cent for moderate contrasts
and within ten per cent even at the highest contrast examined.
Anisotropic spectra introduce directional correlations that are invisible
to a linear approximation; nevertheless, inclusion of the explicit
two-point cumulant at second order recovers the numerical benchmark to
within five per cent across the entire range of contrasts and spectral
exponents studied.  The third-order term proved numerically small in all
cases considered, indicating that the series is effectively converged
once two-point information is accounted for.  An important physical
trend emerging from both theory and simulation is that spatial
fluctuations invariably reduce the effective conductivity relative to
the arithmetic mean, even when highly conducting regions are present
locally; the reduction is most pronounced when long-wavelength
fluctuations are strongly suppressed and when spectral anisotropy is
large.

Because the leading corrections are expressed as simple integrals over
the power spectrum, the present formalism offers a practical route for
inverse material design.  By prescribing target features in
$S_g(\mathbf k)$—for example, suppressing low-frequency modes along one
direction to induce anisotropy—designers can predict, without recourse
to full‐field simulation, how such changes will shift the bulk
conductivity.  This capability slots naturally into computational
materials design platforms that exploit spectral descriptors
\cite{xu2019multiple, xu2022correlation, shi2023computational}, enabling
rapid navigation of large design spaces for architected or graded
composites.  Although the present work focused on thermal or electrical
conduction, the underlying mathematics carries over to other transport
and mechanical properties; extensions to elastic moduli
\cite{torquato1997exact} and frequency-dependent wave phenomena
\cite{torquato2021nonlocal} follow by substituting the appropriate
Green’s function in the starting integral equation.

In closing, the contrast–expansion theory developed here unifies
higher-order microstructural effects with spectral descriptors in an
analytical, computationally efficient framework.  Its accuracy across a
broad palette of isotropic and anisotropic microstructures, together
with its straightforward extensibility, makes it a valuable tool for
both fundamental studies of transport in disordered media and the
inverse design of materials with tailored, directionally dependent
properties.




\smallskip

\bibliography{main}

\end{document}